\batchmode
\makeatletter
\def\input@path{{\string"D:/Rushi/Dropbox/Rushi/Research/Repo/tex/LaTeX/SCC/publications/2017 ACC/Double Integral Concurrent Learning/\string"}}
\makeatother
\documentclass[conference, twocolumn]{IEEEConf}
\usepackage[T1]{fontenc}
\usepackage[latin9]{inputenc}
\usepackage{array}
\usepackage{verbatim}
\usepackage{booktabs}
\usepackage{units}
\usepackage{mathrsfs}
\usepackage{amsmath}
\usepackage{amsthm}
\usepackage{amssymb}
\usepackage{stmaryrd}
\usepackage{graphicx}
\usepackage{esint}
\usepackage{xargs}[2008/03/08]

\makeatletter

\providecommand{\tabularnewline}{\\}

  \theoremstyle{definition}
  \newtheorem{defn}{\protect\definitionname}
\theoremstyle{definition}
\newtheorem{assumption}{Assumption}
  \theoremstyle{plain}
  \newtheorem{lem}{\protect\lemmaname}
  \theoremstyle{plain}
  \newtheorem{thm}{\protect\theoremname}

\usepackage{cite}
\usepackage{nicefrac}
\usepackage{slashbox}
\IEEEoverridecommandlockouts

\makeatother

\providecommand{\definitionname}{Definition}
\providecommand{\lemmaname}{Lemma}
\providecommand{\theoremname}{Theorem}

\begin{document}

\title{Online Output-Feedback Parameter and State Estimation for Second
Order Linear Systems\thanks{Rushikesh Kamalapurkar is with the School of Mechanical and Aerospace
Engineering, Oklahoma State University, Stillwater, OK. Email: rushikesh.kamalapurkar@okstate.edu}}

\author{Rushikesh Kamalapurkar}
\maketitle
\begin{abstract}
In this paper, a concurrent learning based adaptive observer is developed
for a class of second-order linear time-invariant systems with uncertain
system matrices. The developed technique yields an exponentially convergent
state estimator and an exponentially convergent parameter estimator.
As opposed to \textit{persistent} excitation required for parameter
convergence in traditional adaptive methods, excitation over a finite
time-interval is sufficient for the developed technique to achieve
exponential convergence. Simulation results in both noise-free and
noisy environments are presented to validate the design.
\end{abstract}

\section{Introduction}

\global\long\def\d{\textnormal{d}}

\global\long\def\i{\textnormal{i}}

\global\long\def\sgn{\operatorname{sgn}}

\global\long\def\proj{\operatornamewithlimits{proj}}

\global\long\def\argmin{\operatornamewithlimits{arg\:min}}

\global\long\def\r{\mathbb{R}}

\global\long\def\tr{\operatorname{tr}}

\global\long\def\Sgn{\operatorname{SGN}}

\global\long\def\k{\operatorname{K}}

\global\long\def\co{\operatorname{co}}

\global\long\def\ae{\operatorname{a.e.}}

\global\long\def\b{\operatorname{B}}

\global\long\def\n{\mathbb{N}}

\global\long\def\diag{\operatorname{diag}}

\global\long\def\rank{\operatorname{rank}}

\global\long\def\ind{\operatorname{\mathbf{1}}}

\newcommandx\fid[5][usedefault, addprefix=\global, 1=, 2=, 3=, 4=, 5=]{\tensor*[_{#3}^{#2}]{#1}{_{#5}^{#4}}}

\global\long\def\g{\operatorname{g}}

\global\long\def\eval#1{\left.#1\right|}

\global\long\def\lab#1#2{\underset{#2}{\underbrace{#1}}}

\global\long\def\zero{\operatorname{0}}

\global\long\def\e{\textnormal{e}}

\global\long\def\vecop{\operatorname{vec}}

\global\long\def\id{\operatorname{I}}

\global\long\def\re{\operatorname{re}}

\global\long\def\im{\operatorname{im}}

\global\long\def\lap#1{\mathscr{L}\left\{  #1\right\}  }

Over
the the past decade, the role of autonomy in everyday life has seen
an unprecedented growth. As a result, the tasks performed by autonomous
systems have also grown in complexity. Adaptive control methods have
emerged as a tool to address a subset of the challenges posed by complexity.
In particular, autonomous systems typically operate in uncertain and
changing environments. The ability to learn the uncertainty and to
adapt to changes is thus an integral part of a modern control system.

Traditional adaptive control methods handle uncertainty in the system
dynamics by maintaining a parametric estimate of the model and utilizing
it to generate a feedforward control signal \cite{SCC.Ioannou1996,SCC.Sastry1989a,SCC.Krstic1995}.
While the feedforward-feedback architecture guarantees stability of
the closed-loop, the control law is not robust to disturbances, and
seldom provides information regarding the quality of the estimated
model \cite{SCC.Ioannou1996,SCC.Sastry1989a}. In addition to system
identification, parameter convergence in adaptive control schemes
provides increased robustness and improved transient performance (cf.
\cite{SCC.Duarte.Narendra1989,SCC.Krstic.Kokotovic.ea1993,SCC.Chowdhary.Johnson2011a}).
Modifications such as $\sigma-$modification (cf. \cite[Section 8.4.1]{SCC.Ioannou1996})
and $e-$modification (cf. \cite{SCC.Narendra1987}) result in robust
adaptive controllers, however, the parameter estimates generally do
not converge to the true values of the corresponding parameters \cite{SCC.Narendra.Annaswamy1986,SCC.Narendra1987,SCC.Sastry1989a,SCC.Ioannou1996,SCC.Volyanskyy.Calise.ea2006}.
The parameters can be shown to converge under persistent excitation;
however, in addition to the control effort required to maintain excitation,
persistent excitation can lead to mechanical fatigue, and often directly
conflicts with control objectives such as regulation and tracking.

Recently, a novel data-driven concurrent learning (CL) adaptive control
method that achieves parameter convergence under a finite excitation
condition was developed in results such as \cite{SCC.Chowdhary.Johnson2011a,SCC.Chowdhary.Yucelen.ea2012,SCC.Kersting.Buss2014}.
In CL adaptive control, parameter convergence is achieved by storing
data during time-intervals when the system is excited, and then utilizing
the stored data to drive adaptation when excitation is unavailable.
Since excitation is required only over a finite time-interval, energy
utilization and mechanical fatigue can be kept to a minimum, and asymptotic
objectives such as regulation and tracking can be effectively achieved.
Furthermore, CL adaptive control methods possess similar robustness
to bounded disturbances as $\sigma-$modification, $e-$modification,
etc, without the associated drawbacks such as drawing the parameter
estimates to arbitrary set-points \cite{SCC.Chowdhary.Johnson2011a,SCC.Chowdhary.Yucelen.ea2012,SCC.Chowdhary.Muehlegg.ea2013,SCC.Kersting.Buss2014}.

Adaptation techniques similar to the CL method were utilized to implement
reinforcement learning under finite excitation conditions in results
such as \cite{SCC.Modares.Lewis.ea2014,SCC.Kamalapurkar.Klotz.ea2014a,SCC.Luo.Wu.ea2014,SCC.Kamalapurkar.Walters.ea2016,SCC.Bian.Jiang2016,SCC.Kamalapurkar.Rosenfeld.eatoappear}.
CL methods have also been extended to classes of switched systems
(cf. \cite{SCC.DeLaTorre.Chowdhary.ea2013}) and systems driven by
stochastic processes (cf. \cite{SCC.Chowdhary.Kingravi.ea2015}).
A major drawback of CL methods is that they require numerical differentiation
of the state measurements. CL methods that do not require numerical
differentiation of the state measurements are developed in results
such as \cite{arXivKamalapurkar.Reish.ea2015} and \cite{ParikhKamalapurkarDixonarXiv:1512.03464},
however, they require full state feedback. Since full state feedback
is often not available, the development of an output-feedback CL framework
is well-motivated.

In this paper, a CL-based adaptive observer is developed for a class
of second-order linear time-invariant systems. The elements of the
system matrices are assumed to be uncertain and the dimensions of
the matrices are assumed to be known. The developed technique yields
an exponentially convergent state estimator and an exponentially convergent
parameter estimator. Excitation over a finite time-interval (as opposed
to \textit{persistent} excitation) is required for exponential convergence.
Simulation results are provided in a noise-free environment to validate
the design. Simulation results with added measurement noise are also
provided to demonstrate robustness to sensor noise.

In the following, a linear error system is developed in Section \ref{sec:Error-System-for}
to facilitate CL-based adaptation. A CL-based parameter estimator
is designed in Section \ref{sec:Parameter-Estimator-Design}. A state-observer
that utilizes the parameter estimates to estimate the generalized
velocity is developed in Section \ref{sec:State-Observer-Design}.
A Lyapunov-based stability analysis of the parameter estimator and
the state observer is presented in Section \ref{sec:Stability-Analysis}.
Section \ref{sec:Simulations} presents numerical simulation results,
and Section \ref{sec:Conclusion} presents concluding remarks and
a few comments on possible extensions of the developed technique.

\section{\label{sec:Error-System-for}Error System for Estimation}

Consider a second order linear system of the form
\begin{align}
\dot{p}\left(t\right) & =q\left(t\right),\nonumber \\
\dot{q}\left(t\right) & =Ax\left(t\right)+Bu\left(t\right),\nonumber \\
y\left(t\right) & =p\left(t\right),\label{eq:Linear System}
\end{align}
where $p:\mathbb{R}_{\geq t_{0}}\to\r^{n}$ and $q:\r_{\geq t_{0}}\to\r^{n}$
denote the generalized position states and the generalized velocity
states, respectively, $x\triangleq\begin{bmatrix}p^{T} & q^{T}\end{bmatrix}^{T}$
is the system state, $A\in\r^{n\times2n}$ and $B\in\r^{n\times m}$
denote the system matrices, and $y:\r_{\geq t_{0}}\to\r^{n}$ denotes
the output. The objective is to design an adaptive estimator to identify
the unknown matrices $A$ and $B$, online, using input-output measurements.
It is assumed that the system is controlled using a stabilizing input,
i.e., $x,\:u\in\mathcal{L}_{\infty}$. Systems of the form (\ref{eq:Linear System})
can be obtained through linearization of second-order Euler-Lagrange
models, and hence, represent a wide class of physical plants, including
but not limited to robotic manipulators and autonomous ground, aerial,
and underwater vehicles.

To obtain an error signal for parameter identification, the system
in (\ref{eq:Linear System}) is expressed in the form
\begin{equation}
\dot{q}\left(t\right)=A_{1}p\left(t\right)+A_{2}q\left(t\right)+Bu\left(t\right),\label{eq:Pre Integral Form}
\end{equation}
where $A_{1}\in\r^{n\times n}$ and $A_{2}\in\r^{n\times n}$ are
constant matrices such that $A=\begin{bmatrix}A_{1} & A_{2}\end{bmatrix}$.
Integrating (\ref{eq:Pre Integral Form}) over the interval $\left[t-T_{1},t\right]$
for some constant $T_{1}\in\r_{>0}$,
\begin{multline}
q\left(t\right)-q\left(t-T_{1}\right)=A_{1}\intop_{t-T_{1}}^{t}p\left(\tau\right)\d\tau+A_{2}\intop_{t-T_{1}}^{t}q\left(\tau\right)\d\tau\\
+B\intop_{t-T_{1}}^{t}u\left(\tau\right)\d\tau.\label{eq:Single Integral Form}
\end{multline}
Integrating again over the interval $\left[t-T_{2},t\right]$ for
some constant $T_{2}\in\r_{>0}$,
\begin{multline}
\intop_{t-T_{2}}^{t}\left(q\left(\sigma\right)-q\left(\sigma-T_{1}\right)\right)\d\sigma=A_{1}\intop_{t-T_{2}}^{t}\intop_{\sigma-T_{1}}^{\sigma}p\left(\tau\right)\d\tau\d\sigma\\
+A_{2}\intop_{t-T_{2}}^{t}\intop_{\sigma-T_{1}}^{\sigma}q\left(\tau\right)\d\tau\d\sigma+B\intop_{t-T_{2}}^{t}\intop_{\sigma-T_{1}}^{\sigma}u\left(\tau\right)\d\tau\d\sigma.\label{eq:Double Integral Form}
\end{multline}
Using the Fundamental Theorem of Calculus and the fact that $q\left(t\right)=\dot{p}\left(t\right)$,
\begin{multline}
p\left(t\right)-p\left(t-T_{2}\right)-p\left(t-T_{1}\right)+p\left(t-T_{2}-T_{1}\right)=\\
A_{1}F\left(t\right)+A_{2}G\left(t\right)+BU\left(t\right).\label{eq:Derivative Free Form}
\end{multline}
where 
\begin{equation}
F\left(t\right)\triangleq\begin{cases}
\intop_{t-T_{2}}^{t}\intop_{\sigma-T_{1}}^{\sigma}p\left(\tau\right)\d\tau\d\sigma, & t\in\left[t_{0}+T_{1}+T_{2},\infty\right),\\
0, & t<t_{0}+T_{1}+T_{2},
\end{cases}\label{eq:F}
\end{equation}
 
\begin{equation}
G\left(t\right)\!\triangleq\!\begin{cases}
\!\intop_{t-T_{2}}^{t}\!\!\left(p\left(\!\sigma\!\right)\!-\!p\left(\sigma\!-\!T_{1}\right)\!\right)\d\sigma, & t\!\in\!\left[t_{0}\!+\!T_{1}\!+\!T_{2},\infty\!\right),\\
0 & t<t_{0}+T_{1}+T_{2},
\end{cases}\label{eq:G}
\end{equation}
and 
\begin{equation}
U\left(t\right)\triangleq\begin{cases}
\intop_{t-T_{2}}^{t}\intop_{\sigma-T_{1}}^{\sigma}u\left(\tau\right)\d\tau\d\sigma, & t\in\left[t_{0}+T_{1}+T_{2},\infty\right),\\
0 & t<t_{0}+T_{1}+T_{2}.
\end{cases}\label{eq:U}
\end{equation}
The utility of the integral form in (\ref{eq:Derivative Free Form})
is that it is independent of the generalized velocity states, $q$.
The expression in (\ref{eq:Derivative Free Form}) can be rearranged
to form the linear error system
\begin{equation}
\mathcal{F}\left(t\right)=\mathcal{G}\left(t\right)\theta,\:\forall t\in\r_{\geq t_{0}}.\label{eq:Linear Error System}
\end{equation}
In (\ref{eq:Linear Error System}), $\theta$ is a vector of unknown
parameters, defined as $\theta\triangleq\begin{bmatrix}\vecop\left(A_{1}\right)^{T} & \vecop\left(A_{2}\right)^{T} & \vecop\left(B\right)^{T}\end{bmatrix}^{T}\in\r^{2n^{2}+mn}$,
where $\vecop\left(\cdot\right)$ denotes the vectorization operator
and the matrices $\mathcal{F}:\r_{\geq0}\to\r^{n}$ and $\mathcal{G}:\r_{\geq0}\to\r^{n\times2n^{2}+mn}$
are defined as 
\begin{align*}
\mathcal{F}\left(t\right) & \triangleq\begin{cases}
\begin{gathered}p\left(t\!-\!T_{2}\!-\!T_{1}\right)\!-\!p\left(t\!-\!T_{1}\right)\\
\!+p\left(t\right)\!-\!p\left(t\!-\!T_{2}\right),
\end{gathered}
 & t\!\in\!\left[t_{0}\!+\!T_{1}\!+\!T_{2},\infty\right),\\
0 & t<t_{0}+T_{1}+T_{2}.
\end{cases}\\
\mathcal{G}\left(t\right) & \triangleq\begin{bmatrix}\left(F\left(t\right)\varotimes\id_{n}\right)^{T} & \left(G\left(t\right)\varotimes\id_{n}\right)^{T} & \left(U\left(t\right)\varotimes\id_{n}\right)^{T}\end{bmatrix},
\end{align*}
where $\id_{n}$ denotes an $n\times n$ identity matrix, and $\varotimes$
denotes the Kronecker product. Note that even though the linear relationship
in (\ref{eq:Linear Error System}) is valid for all $t\in\r_{\geq t_{0}},$
it provides useful information about the vector $\theta$ only after
$t\geq t_{0}+T_{1}+T_{2}$. 

The linear error system in (\ref{eq:Linear Error System}) motivates
the adaptive estimation scheme that follows. The design is inspired
by the \textit{concurrent learning} (cf. \cite{SCC.Chowdhary2010})
technique. Concurrent learning enables parameter convergence in adaptive
control by using stored data to update the parameter estimates. Traditionally,
adaptive control methods guarantee parameter convergence only if the
appropriate PE conditions are met (cf. \cite[Chapter 4]{SCC.Ioannou1996}).
Concurrent learning uses stored data to soften the PE condition to
an excitation condition over a finite time-interval. Concurrent learning
methods such as \cite{SCC.Chowdhary.Johnson2011a} and \cite{SCC.Kersting.Buss2014}
require numerical differentiation of the system state, and concurrent
learning techniques such as \cite{ParikhKamalapurkarDixonarXiv:1512.03464}
and \cite{arXivKamalapurkar.Reish.ea2015} require full state measurements.
In the following, a concurrent learning method that utilizes only
the output measurements is developed. 

\section{\label{sec:Parameter-Estimator-Design}Parameter Estimator Design}

To obtain output-feedback concurrent learning update law for the parameter
estimates, a history stack denoted by $\mathcal{H}$ is utilized.
The history stack is a set of ordered pairs $\left\{ \left(\mathcal{F}_{i},\mathcal{G}_{i}\right)\right\} _{i=1}^{M}$
such that 
\begin{equation}
\mathcal{F}_{i}=\mathcal{G}_{i}\theta,\:\forall i\in\left\{ 1,\cdots,M\right\} .\label{eq:History Stack Compatibility}
\end{equation}
If a history stack that satisfies (\ref{eq:Online Recording}) is
not available a priori, it can be recorded online, using the relationship
in (\ref{eq:Linear Error System}), by selecting a set of time-instances
$\left\{ t_{i}\right\} _{i=1}^{M}$ and letting 
\begin{align}
\mathcal{F}_{i} & =\mathcal{F}\left(t_{i}\right),\nonumber \\
\mathcal{G}_{i} & =\mathcal{G}\left(t_{i}\right).\label{eq:Online Recording}
\end{align}
Furthermore, a singular value maximization algorithm is used to select
the time instances $\left\{ t_{i}\right\} _{i=1}^{M}$. That is, a
data-point $\left\{ \left(\mathcal{F}_{j},\mathcal{G}_{j}\right)\right\} $
in the history stack is replaced by a new data-point $\left\{ \left(\mathcal{F}^{*},\mathcal{G}^{*}\right)\right\} $,
where $\mathcal{F}^{*}=\mathcal{F}\left(t\right)$ and $\mathcal{G}^{*}=\mathcal{G}\left(t\right)$,
for some $t$, only if 
\[
\lambda_{\min}\!\left\{ \!\sum_{i\neq j}\mathcal{G}_{i}^{T}\mathcal{G}_{i}\!+\!\mathcal{G}_{j}^{T}\mathcal{G}_{j}\!\right\} \!<\!\lambda_{\min}\!\left\{ \!\sum_{i\neq j}\mathcal{G}_{i}^{T}\mathcal{G}_{i}\!+\!\mathcal{G}^{*T}\mathcal{G}^{*}\!\right\} ,
\]
where $\lambda_{\min}\left\{ \cdot\right\} $ denotes the minimum
Eigenvalue of a matrix. 
\begin{defn}
A history stack $\left\{ \left(\mathcal{F}_{i},\mathcal{G}_{i}\right)\right\} _{i=1}^{M}$
is called \textit{full rank }if there exists a constant $\underline{c}\in\r$
such that 
\begin{equation}
0<\underline{c}<\lambda_{\min}\left\{ \mathscr{G}\right\} ,\label{eq:Rank Condition}
\end{equation}
where the matrix $\mathscr{G}\in\r^{\left(2n^{2}+mn\right)\times\left(2n^{2}+mn\right)}$
is defined as $\mathscr{G}\triangleq\sum_{i=1}^{M}\mathcal{G}_{i}^{T}\mathcal{G}_{i}$.
\end{defn}
The concurrent learning update law to estimate the unknown parameters
is then given by%
\begin{equation}
\dot{\hat{\theta}}\left(t\right)=k_{\theta}\Gamma\left(t\right)\sum_{i=1}^{M}\mathcal{G}_{i}^{T}\left(\mathcal{F}_{i}-\mathcal{G}_{i}\hat{\theta}\left(t\right)\right),\label{eq:Theta Dynamics}
\end{equation}
where $k_{\theta}\in\r_{>0}$ is a constant adaptation gain and $\Gamma:\r_{\geq0}\to\r^{\left(2n^{2}+mn\right)\times\left(2n^{2}+mn\right)}$
is the least-squares gain updated using the update law
\begin{equation}
\dot{\Gamma}\left(t\right)=\beta_{1}\Gamma\left(t\right)-k_{\theta}\Gamma\left(t\right)\sum_{i=1}^{M}\mathcal{G}_{i}^{T}\mathcal{G}_{i}\Gamma\left(t\right).\label{eq:Gamma Dynamics}
\end{equation}
Using arguments similar to Corollary 4.3.2 in \cite{SCC.Ioannou1996},
it can be shown that provided $\lambda_{\min}\left\{ \Gamma^{-1}\left(t_{0}\right)\right\} >0$,
the least squares gain matrix satisfies 
\begin{equation}
\underline{\Gamma}\id_{\left(2n^{2}+mn\right)}\leq\Gamma\left(t\right)\leq\overline{\Gamma}\id_{\left(2n^{2}+mn\right)},\label{eq:StaFGammaBound}
\end{equation}
where $\underline{\Gamma}$ and $\overline{\Gamma}$ are positive
constants, and $\id_{n}$ denotes an $n\times n$ identity matrix.
The following finite-excitation assumption is necessary for the update
law in (\ref{eq:Theta Dynamics}) to result in an exponentially convergent
parameter estimator. 
\begin{assumption}
For a given $M\in\mathbb{N}$ and $\underline{c}\in\r_{>0}$, there
exists a set of time instances $\left\{ t_{i}\right\} _{i=1}^{M}$
such that a history stack recorded using (\ref{eq:Online Recording})
is full rank.
\end{assumption}
Since the history stack is updated using a singular value maximization
algorithm, the matrix $\mathscr{G}$ is a piece-wise constant function
of time. The use of singular value maximization to update the history
stack implies that once the matrix $\mathscr{G}$ satisfies (\ref{eq:Rank Condition}),
at some $t=T$, and for some $\underline{c}$, the condition $\underline{c}<\lambda_{\min}\left\{ \mathscr{G}\left(t\right)\right\} $
holds for all $t\geq T$. The following section details the design
of an exponentially convergent adaptive state-observer.

\section{\label{sec:State-Observer-Design}State Observer Design}

To facilitate parameter estimation based on a prediction error, a
state observer is developed in the following. To facilitate the design,
the dynamics in (\ref{eq:Linear System}) are expressed in the form
\begin{align*}
\dot{p}\left(t\right) & =q\left(t\right),\\
\dot{q}\left(t\right) & =Y\left(x\left(t\right),u\left(t\right)\right)\theta,
\end{align*}
where $Y:\r^{n}\times\r^{m}\to\r^{n\times\left(2n^{2}+mn\right)}$
is defined as $Y\left(x,u\right)=\begin{bmatrix}\left(p\varotimes\id_{n}\right)^{T} & \left(q\varotimes\id_{n}\right)^{T} & \left(u\varotimes\id_{n}\right)^{T}\end{bmatrix}.$
The adaptive state observer is then designed as
\begin{align}
\dot{\hat{p}}\left(t\right) & =\hat{q}\left(t\right),\quad\hat{p}\left(t_{0}\right)=p\left(t_{0}\right),\nonumber \\
\dot{\hat{q}}\left(t\right) & =Y\left(x\left(t\right),u\left(t\right)\right)\hat{\theta}\left(t\right)+\nu\left(t\right),\quad\hat{q}\left(t_{0}\right)=0,\label{eq:State observer}
\end{align}
where $\hat{p}:\r_{\geq t_{0}}\to\r^{n}$, $\hat{q}:\r_{\geq t_{0}}\to\r^{n}$,
$\hat{x}:\r_{\geq t_{0}}\to\r^{n}$, and $\hat{\theta}:\r_{\geq t_{0}}\to\r^{n}$
are estimates of $p$, $q$, $x$, and $\theta$, respectively, $\nu$
is the feedback component of the identifier, to be designed later,
and the prediction error $\tilde{p}:\r_{\geq t_{0}}\to\r^{n}$ is
defined as 
\[
\tilde{p}\left(t\right)=p\left(t\right)-\hat{p}\left(t\right).
\]

The update law for the generalized velocity estimate depends on the
entire state $x$. However, using the structure of the matrix $Y$
and integrating by parts, the observer can be implemented without
using generalized velocity measurements. Consider the integral form
of (\ref{eq:State observer})
\[
\hat{q}\left(t\right)-\hat{q}\left(t_{0}\right)=\intop_{t_{0}}^{t}\left(Y\left(x\left(\tau\right),u\left(\tau\right)\right)\hat{\theta}\left(\tau\right)+\nu\left(\tau\right)\right)\d\tau.
\]
Using the definition of $Y$ and $\theta$, and expanding the integral,
\begin{multline*}
\hat{q}\left(t\right)-\hat{q}\left(t_{0}\right)=\intop_{t_{0}}^{t}\left(p\left(\tau\right)\varotimes\id_{n}\right)^{T}\vecop\left(\hat{A}_{1}\left(\tau\right)\right)\d\tau\\
+\intop_{t_{0}}^{t}\left(u\left(\tau\right)\varotimes\id_{n}\right)^{T}\vecop\left(\hat{B}\left(\tau\right)\right)\d\tau+\intop_{t_{0}}^{t}\nu\left(\tau\right)\d\tau\\
+\intop_{t_{0}}^{t}\left(q\left(\tau\right)\varotimes\id_{n}\right)^{T}\vecop\left(\hat{A}_{2}\left(\tau\right)\right)\d\tau.
\end{multline*}
The last term of the integral can be further expanded using integration
by parts to yield
\begin{multline*}
\intop_{t_{0}}^{t}\left(q\left(\tau\right)\varotimes\id_{n}\right)^{T}\vecop\left(\hat{A}_{2}\left(\tau\right)\right)\d\tau=\\
\left(p\left(t\right)\varotimes\id_{n}\right)^{T}\vecop\left(\hat{A}_{2}\left(t\right)\right)-\left(p\left(t_{0}\right)\varotimes\id_{n}\right)^{T}\vecop\left(\hat{A}_{2}\left(t_{0}\right)\right)\\
-\intop_{t_{0}}^{t}\left(p\left(\tau\right)\varotimes\id_{n}\right)^{T}\vecop\left(\dot{\hat{A}}_{2}\left(\tau\right)\right)\d\tau.
\end{multline*}
Thus, the update law in (\ref{eq:State observer}) can be implemented
without generalized velocity measurements as
\begin{multline}
\hat{q}\left(t\right)=\intop_{t_{0}}^{t}\left(u\left(\tau\right)\varotimes\id_{n}\right)^{T}\vecop\left(\hat{B}\left(\tau\right)\right)\d\tau+\intop_{t_{0}}^{t}\nu\left(\tau\right)\d\tau\\
+\hat{q}\left(t_{0}\right)+\intop_{t_{0}}^{t}\!\!\left(p\left(\tau\right)\varotimes\id_{n}\right)^{T}\!\!\left(\!\vecop\left(\hat{A}_{1}\left(\tau\right)\right)\!-\!\vecop\left(\dot{\hat{A}}_{2}\left(\tau\right)\right)\!\right)\d\tau\\
+\left(p\left(t\right)\varotimes\id_{n}\right)^{T}\vecop\left(\hat{A}_{2}\left(t\right)\right)-\left(p\left(t_{0}\right)\varotimes\id_{n}\right)^{T}\vecop\left(\hat{A}_{2}\left(t_{0}\right)\right)\label{eq:IntegralUpdate}
\end{multline}

To facilitate the design of the feedback component $\nu$, let 
\begin{equation}
r\left(t\right)=\tilde{q}\left(t\right)+\alpha\tilde{p}\left(t\right)+\eta\left(t\right),\label{eq:r}
\end{equation}
where the signal $\eta$ is added to compensate for the fact that
the generalized velocity state, $q$, is not measurable. Based on
the subsequent stability analysis, the signal $\eta$ is designed
as the output of the dynamic filter 
\begin{align}
\dot{\eta}\left(t\right) & =-\beta\eta\left(t\right)-kr\left(t\right)-\alpha\tilde{q}\left(t\right),\quad\eta\left(t_{0}\right)=0,\label{eq:eta Update}
\end{align}
and the feedback component $\nu$ is designed as 
\begin{equation}
\nu\left(t\right)=\tilde{p}\left(t\right)-\left(k+\alpha+\beta\right)\eta\left(t\right).\label{eq:Feedback}
\end{equation}
The design of the signals $\eta$ and $\nu$ to estimate the state
from output measurements is inspired by the $p-$filter (cf. \cite{SCC.Xian2004c}).
Similar to the update law for the generalized velocity, using the
the fact that $\tilde{p}\left(t_{0}\right)=0$, the signal $\eta$
can be implemented using the integral form
\begin{equation}
\eta\left(t\right)=-\intop_{t_{0}}^{t}\left(\beta+k\right)\eta\left(\tau\right)\d\tau-\intop_{t_{0}}^{t}k\alpha\tilde{p}\left(\tau\right)\d\tau-\left(k+\alpha\right)\tilde{p}\left(t\right).\label{eq:IntegralUpdateEta}
\end{equation}
A Lyapunov-based analysis of the parameter and the state estimation
errors is presented in the following section.

\section{\label{sec:Stability-Analysis}Stability Analysis}

To facilitate the analysis, (\ref{eq:History Stack Compatibility})
and (\ref{eq:Theta Dynamics}) are used to express the dynamics of
the parameter estimation error as
\begin{equation}
\dot{\tilde{\theta}}\left(t\right)=-k_{\theta}\Gamma\left(t\right)\mathscr{G}\left(t\right)\tilde{\theta}\left(t\right).\label{eq:Theta Error Dynamics}
\end{equation}
Since the function $\mathscr{G}:\r_{\geq t_{0}}\to\r^{\left(2n^{2}+mn\right)\times\left(2n^{2}+mn\right)}$
is piece-wise continuous, the trajectories of (\ref{eq:Theta Error Dynamics}),
and of all the subsequent error systems involving $\mathscr{G}$,
are defined in the sense of Carath\'{e}odory. Using the dynamics
in (\ref{eq:Linear System}), (\ref{eq:State observer}), (\ref{eq:eta Update}),
and the design of the feedback component in (\ref{eq:Feedback}),
the time-derivative of the error signal $r$ is given by 
\[
\dot{r}\left(t\right)=Y\left(x\left(t\right),u\left(t\right)\right)\tilde{\theta}\left(t\right)-\tilde{p}\left(t\right)+\left(k+\alpha\right)\eta\left(t\right)-kr\left(t\right).
\]
The analysis is carried out separately over the time intervals $t\in\left[t_{0},t_{0}+t_{M}\right]$
and $t\in\r_{\geq t_{M}}$. It is established that the error trajectories
remain bounded for $t\in\left[t_{0},t_{0}+t_{M}\right]$ and that
the error trajectories decay exponentially to zero for $t\in\r_{\geq t_{M}}$.
The following Lemma establishes boundedness of the parameter estimation
error vector for all $t\in\r_{\geq t_{0}}$.
\begin{lem}
\label{lem:Theta-tilde Bounded}The parameter estimation error vector
satisfies the bound 
\begin{equation}
\left\Vert \tilde{\theta}\left(t\right)\right\Vert \leq\overline{\theta},\:\forall t\in\r_{\geq t_{0}},\label{eq:Theta Bound}
\end{equation}
where $\overline{\theta}\in\r$ is a positive constant.
\end{lem}
\begin{IEEEproof}
The candidate Lyapunov function $V_{\theta}\left(\tilde{\theta},t\right)\triangleq\frac{1}{2}\tilde{\theta}^{T}\Gamma^{-1}\left(t\right)\tilde{\theta}$
can be differentiated along the trajectories of (\ref{eq:Theta Error Dynamics})
and (\ref{eq:Gamma Dynamics}) to yield%
\[
\dot{V}_{\theta}\left(\tilde{\theta}\left(t\right),t\right)\leq-\frac{k_{\theta}}{2}\tilde{\theta}^{T}\left(t\right)\mathscr{G}\left(t\right)\tilde{\theta}\left(t\right)-\frac{\beta_{1}}{2}\tilde{\theta}^{T}\left(t\right)\Gamma^{-1}\left(t\right)\tilde{\theta}\left(t\right).
\]
The bound in (\ref{eq:StaFGammaBound}) yields
\[
\dot{V}_{\theta}\left(\tilde{\theta}\left(t\right),t\right)\leq-\frac{k_{\theta}}{2}\tilde{\theta}^{T}\left(t\right)\mathscr{G}\left(t\right)\tilde{\theta}\left(t\right).
\]
Since $\mathscr{G}\left(t\right)$ is a positive semidefinite matrix
for all $t\in\r_{\geq t_{0}}$, the candidate Lyapunov function satisfies
\[
V_{\theta}\left(\tilde{\theta}\left(t\right)\right)\leq\overline{V}_{\theta},\:\forall t\in\r_{\geq t_{0}},
\]
where $\overline{V}_{\theta}\triangleq V_{\theta}\left(\tilde{\theta}\left(t_{0}\right),t_{0}\right)$.
Using the fact that $\underline{\gamma}\left\Vert \tilde{\theta}\right\Vert ^{2}\leq V_{\theta}\left(\tilde{\theta},t\right)$,
for all $\left(\tilde{\theta},t\right)\in\r^{\left(2n^{2}+mn\right)}\times\r_{\geq t_{0}}$,
where $\underline{\gamma}\triangleq\nicefrac{1}{2\overline{\Gamma}}$,
it is concluded that the parameter estimation error satisfies (\ref{eq:Theta Bound}).
\end{IEEEproof}
For brevity of notation, time-dependence of all the signals is suppressed
hereafter. The following Lemma establishes boundedness of the observer
error signals for all $t\in\r_{\geq t_{0}}$.
\begin{lem}
\label{lem:Errors Bounded}Provided the observer gains are selected
such that 
\[
\beta>\frac{\left(1+\alpha^{2}\right)^{2}}{4\alpha},
\]
the state-estimation error, $\tilde{x}$, and the auxiliary observer
error signals, $\eta$ and $r$, are bounded for all $t\in\r_{\geq t_{0}}$.
\end{lem}
\begin{IEEEproof}
To establish boundedness of the observer error signals, consider the
candidate Lyapunov function
\begin{equation}
V_{r}\left(\tilde{p},r,\eta\right)\triangleq\frac{1}{2}\tilde{p}^{T}\tilde{p}+\frac{1}{2}\eta^{T}\eta+\frac{1}{2}r^{T}r.\label{eq:candidate lyapunov 1}
\end{equation}
The time-derivative of (\ref{eq:candidate lyapunov 1}) along the
trajectories of (\ref{eq:Linear System}), (\ref{eq:State observer}),
and (\ref{eq:eta Update}) is given by
\begin{multline*}
\dot{V}_{r}=\tilde{p}^{T}\tilde{q}+\eta^{T}\left(-\beta\eta-kr-\alpha\tilde{q}\right)\\
+r^{T}\left(Y\left(x,u\right)\tilde{\theta}-\tilde{p}+\left(k+\alpha\right)\eta-kr\right).
\end{multline*}
Using (\ref{eq:r}), the Cauchy-Schwartz inequality and simplifying
and canceling common terms, 
\begin{multline*}
\dot{V}_{r}\leq-\begin{bmatrix}\left\Vert \tilde{p}\right\Vert  & \left\Vert \eta\right\Vert \end{bmatrix}\begin{bmatrix}\alpha & \frac{-\left|1-\alpha^{2}\right|}{2}\\
\frac{-\left|1-\alpha^{2}\right|}{2} & \beta-\alpha
\end{bmatrix}\begin{bmatrix}\left\Vert \tilde{p}\right\Vert \\
\left\Vert \eta\right\Vert 
\end{bmatrix}-kr^{T}r\\
+r^{T}Y\left(x,u\right)\tilde{\theta}.
\end{multline*}
Using (\ref{eq:Theta Bound}) and the fact that $x$ and $u$ are
bounded, the matrix $Y$ can be bounded as $\sup_{t\in\r_{\geq t_{0}}}\left\Vert Y\left(x\left(t\right),u\left(t\right)\right)\right\Vert \leq\overline{Y}$
and the derivative of the candidate Lyapunov function can be bounded
as
\begin{multline*}
\dot{V}_{r}\leq-\begin{bmatrix}\left\Vert \tilde{p}\right\Vert  & \left\Vert \eta\right\Vert \end{bmatrix}\begin{bmatrix}\alpha & \frac{-\left|1-\alpha^{2}\right|}{2}\\
\frac{-\left|1-\alpha^{2}\right|}{2} & \beta-\alpha
\end{bmatrix}\begin{bmatrix}\left\Vert \tilde{p}\right\Vert \\
\left\Vert \eta\right\Vert 
\end{bmatrix}-k\left\Vert r\right\Vert ^{2}\\
+\overline{Y}\overline{\theta}\left\Vert r\right\Vert .
\end{multline*}
Completing the squares, using the fact that provided $\beta>\nicefrac{\left(1+\alpha^{2}\right)^{2}}{4\alpha}$,
the matrix 
\[
Q_{r}\triangleq\begin{bmatrix}\alpha & \frac{-\left|1-\alpha^{2}\right|}{2}\\
\frac{-\left|1-\alpha^{2}\right|}{2} & \beta-\alpha
\end{bmatrix}
\]
is positive definite, and letting $\varpi_{r}=\min\left\{ 2\lambda_{\min}\left\{ Q_{r}\right\} ,k\right\} $
\[
\dot{V}_{r}\leq-\varpi_{r}V_{r}+\frac{\overline{Y}^{2}\overline{\theta}^{2}}{2k}.
\]
Hence, the candidate Lyapunov function $V_{r}$ satisfies the bound
$\sup_{t\in\r_{\geq t_{0}}}\left\{ V_{r}\left(\tilde{p}\left(t\right),r\left(t\right),\eta\left(t\right)\right)\right\} \leq\overline{V}_{r}$,
where $\overline{V}_{r}\triangleq\max\left\{ V_{r}\left(t_{0}\right),\nicefrac{\overline{Y}^{2}\overline{\theta}^{2}}{2k\varpi_{r}}\right\} .$
\end{IEEEproof}
In the following, Theorem \ref{thm:Main Thm} demonstrates exponential
convergence of all the error signals to the origin.
\begin{thm}
\label{thm:Main Thm}Provided the hypothesis of Lemma \ref{lem:Errors Bounded}
hold, the learning gains are selected such that 
\[
kk_{\theta}\overline{c}>\frac{\overline{Y}^{2}}{4},
\]
and provided the history stack is populated using the singular value
maximization algorithm, the parameter estimation error, $\tilde{\theta}$,
and the state estimation error, $\tilde{x}$, converge exponentially
to zero.
\end{thm}
\begin{IEEEproof}
Let the candidate Lyapunov function $V:\r^{n}\times\r^{n}\times\r^{n}\times\r^{2n^{2}+mn}\to\r$
be defined as
\begin{equation}
V\left(\tilde{p},r,\eta,\tilde{\theta},t\right)=V_{r}\left(\tilde{p},r,\eta\right)+V_{\theta}\left(\tilde{\theta},t\right).\label{eq:Final Lyapunov Function}
\end{equation}
Consider the time-interval $t\in\left[t_{0},t_{M}\right]$. Lemmas
\ref{lem:Theta-tilde Bounded} and \ref{lem:Errors Bounded} imply
that the candidate Lyapunov function satisfies $V\left(\tilde{p}\left(t\right),r\left(t\right),\eta\left(t\right),\tilde{\theta}\left(t\right),t\right)\leq\overline{V}_{r}+\overline{V}_{\theta}$,
for all $t\in\left[t_{0},t_{M}\right]$. In particular, $V\left(\tilde{p}\left(t_{M}\right),r\left(t_{M}\right),\eta\left(t_{M}\right),\tilde{\theta}\left(t_{M}\right),t_{M}\right)\leq\overline{V}_{r}+\overline{V}_{\theta}$.
Over the time interval $t\in\r_{>t_{M}}$, the time-derivative of
(\ref{eq:Final Lyapunov Function}), along the trajectories of (\ref{eq:Linear System}),
(\ref{eq:State observer}), (\ref{eq:eta Update}), and (\ref{eq:Theta Error Dynamics})
satisfies the bound
\begin{multline}
\dot{V}\leq\tilde{p}^{T}\tilde{q}+\eta^{T}\left(-\beta\eta-kr-\alpha\tilde{q}\right)\\
+r^{T}\left(Y\left(x,u\right)\tilde{\theta}-\tilde{p}+\left(k+\alpha\right)\eta-kr\right)-k_{\theta}\tilde{\theta}^{T}\mathscr{G}\tilde{\theta}.\label{eq:Vdot 1}
\end{multline}
Since the history stack is full rank during the time-interval $t\in\r_{>t_{M}}$,
the matrix $\mathscr{G}$ satisfies the rank condition in (\ref{eq:Rank Condition}).
Hence, (\ref{eq:Vdot 1}) satisfies the bound
\begin{equation}
\dot{V}\leq-\begin{bmatrix}\left\Vert \tilde{p}\right\Vert  & \left\Vert \eta\right\Vert \end{bmatrix}Q_{r}\begin{bmatrix}\left\Vert \tilde{p}\right\Vert \\
\left\Vert \eta\right\Vert 
\end{bmatrix}-\begin{bmatrix}\left\Vert r\right\Vert  & \left\Vert \tilde{\theta}\right\Vert \end{bmatrix}Q_{\theta}\begin{bmatrix}\left\Vert r\right\Vert \\
\left\Vert \tilde{\theta}\right\Vert 
\end{bmatrix},\label{eq:VDot 2}
\end{equation}
where 
\[
Q_{\theta}\triangleq\begin{bmatrix}k & -\frac{\overline{Y}}{2}\\
-\frac{\overline{Y}}{2} & k_{\theta}\underline{c}
\end{bmatrix}.
\]
Provided $\beta>\nicefrac{\left(1+\alpha^{2}\right)^{2}}{4\alpha}$
and $kk_{\theta}\underline{c}>\nicefrac{\overline{Y}^{2}}{4}$, the
matrices $Q_{r}$ and $Q_{\theta}$ are positive definite, and hence,
(\ref{eq:VDot 2}) satisfies the bound
\[
\dot{V}\leq-\varpi V,
\]
where $\varpi\triangleq2\min\left\{ \lambda_{\min}\left\{ Q_{r}\right\} ,\lambda_{\min}\left\{ Q_{\theta}\right\} \right\} $.
Hence, using the Comparison Lemma \cite[Lemma 3.4]{SCC.Khalil2002}
\begin{multline*}
V\left(\tilde{p}\left(t\right),r\left(t\right),\eta\left(t\right),\tilde{\theta}\left(t\right),t\right)\leq\\
V\left(\tilde{p}\left(t_{M}\right),r\left(t_{M}\right),\eta\left(t_{M}\right),\tilde{\theta}\left(t_{M}\right),t_{M}\right)\e^{-\varpi\left(t-t_{M}\right)},
\end{multline*}
$\forall t\in\r_{>t_{M}}$, which implies that 
\[
V\left(\tilde{p}\left(t\right),r\left(t\right),\eta\left(t\right),\tilde{\theta}\left(t\right),t\right)\leq\left(\overline{V}_{\theta}+\overline{V}_{r}\right)\e^{-\varpi\left(t-t_{M}\right)},
\]
$\forall t\in\r_{>t_{M}}$. Hence, the parameter estimation error,
$\tilde{\theta}$, and the state estimation error, $\tilde{x}$, converge
exponentially to the origin.
\end{IEEEproof}

\section{\label{sec:Simulations}Simulations}

\begin{table}
\caption{\label{tab:Simulation-parameters-for}Simulation parameters for the
different simulation runs. The parameters are selected using trial
and error.}

\begin{tabular}{>{\centering}p{0.4\columnwidth}>{\centering}p{0.13\columnwidth}>{\centering}p{0.13\columnwidth}>{\centering}p{0.13\columnwidth}}
\toprule 
 &
\multicolumn{3}{c}{Noise Variance}\tabularnewline
Parameter &
0 &
0.001 &
0.01\tabularnewline
\midrule
$T_{1}$ &
0.5 &
0.9 &
1\tabularnewline
$T_{2}$ &
0.3 &
0.5 &
0.4\tabularnewline
$N$ &
50 &
50 &
150\tabularnewline
$\Gamma\left(t_{0}\right)$ &
$\id_{12}$ &
$\id_{12}$ &
$\id_{12}$\tabularnewline
$\beta_{1}$ &
0.5 &
0.5 &
0.5\tabularnewline
$\alpha$ &
2 &
2 &
2\tabularnewline
$k$ &
10 &
10 &
10\tabularnewline
$\beta$ &
2 &
2 &
2\tabularnewline
$k_{\theta}$ &
\nicefrac{0.5}{N} &
\nicefrac{0.5}{N} &
\nicefrac{0.5}{N}\tabularnewline
\bottomrule
\end{tabular}
\end{table}
\begin{figure}
\includegraphics[width=1\columnwidth]{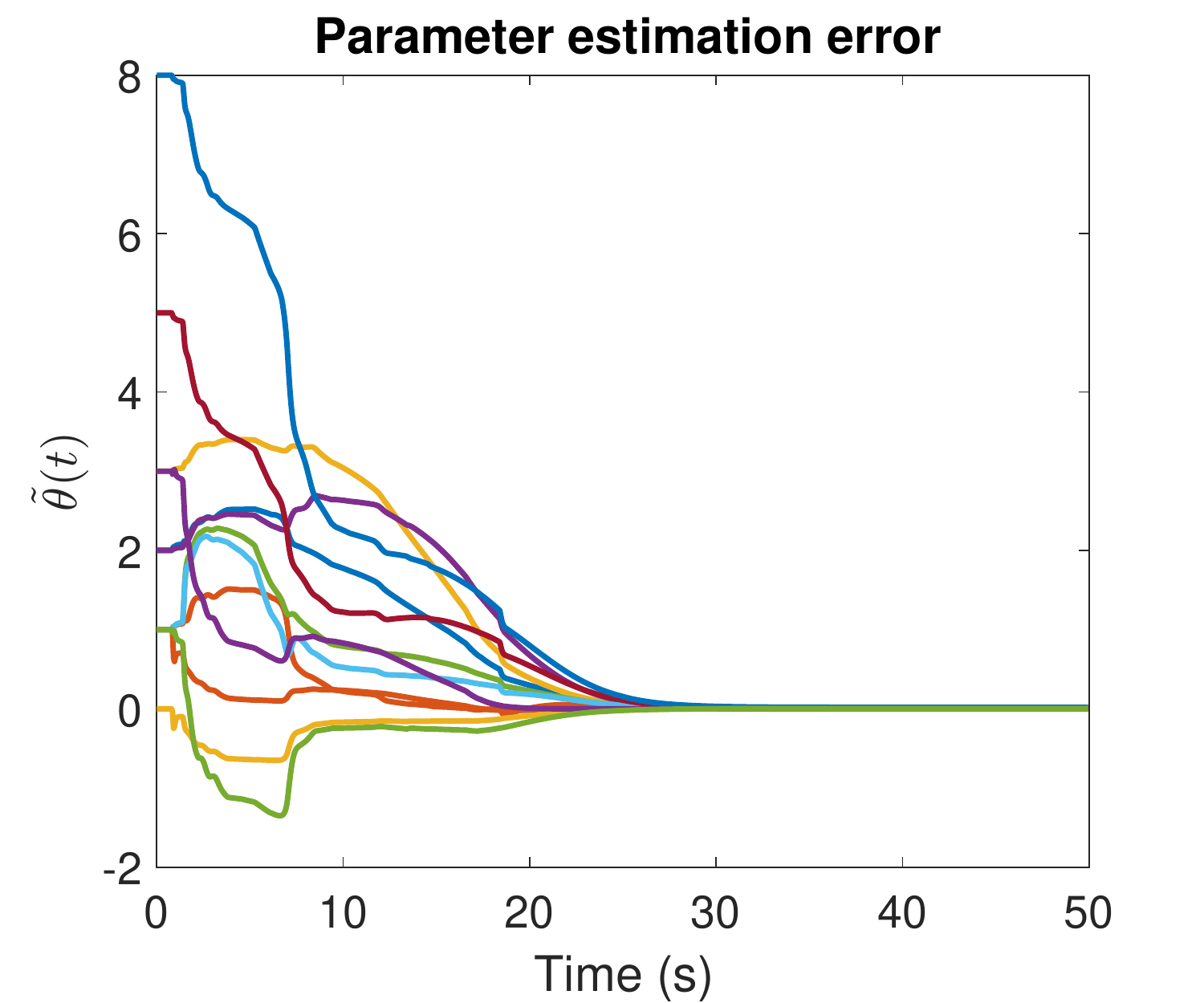}

\caption{\label{fig:ThetaTildeNoNoise}Trajectories of the parameter estimation
errors using noise-free position measurements.}
\end{figure}
\begin{figure}
\includegraphics[width=1\columnwidth]{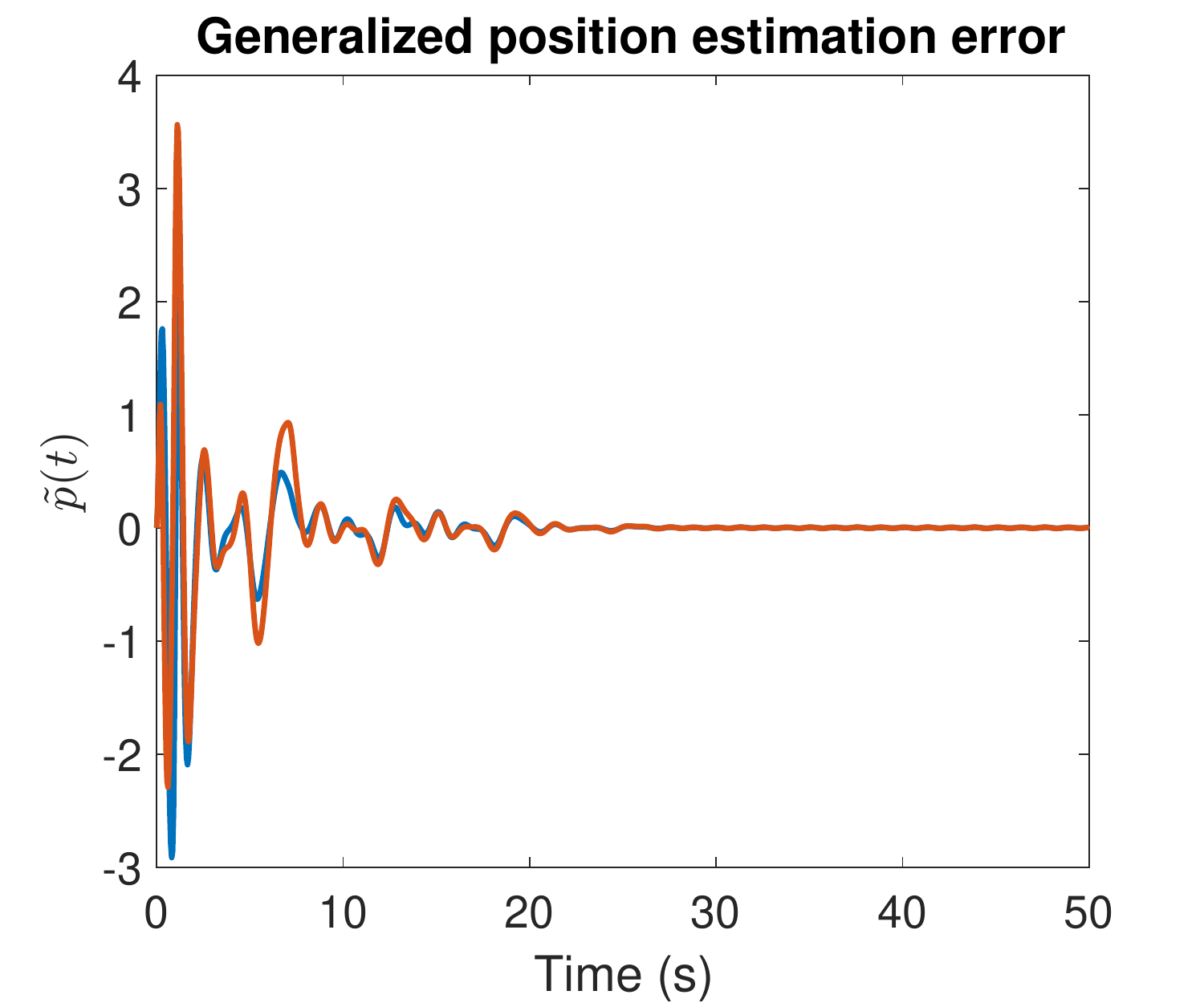}

\caption{\label{fig:pTildeNoNoise}Trajectories of the generalized position
estimation errors using noise-free position measurements.}
\end{figure}
\begin{figure}
\includegraphics[width=1\columnwidth]{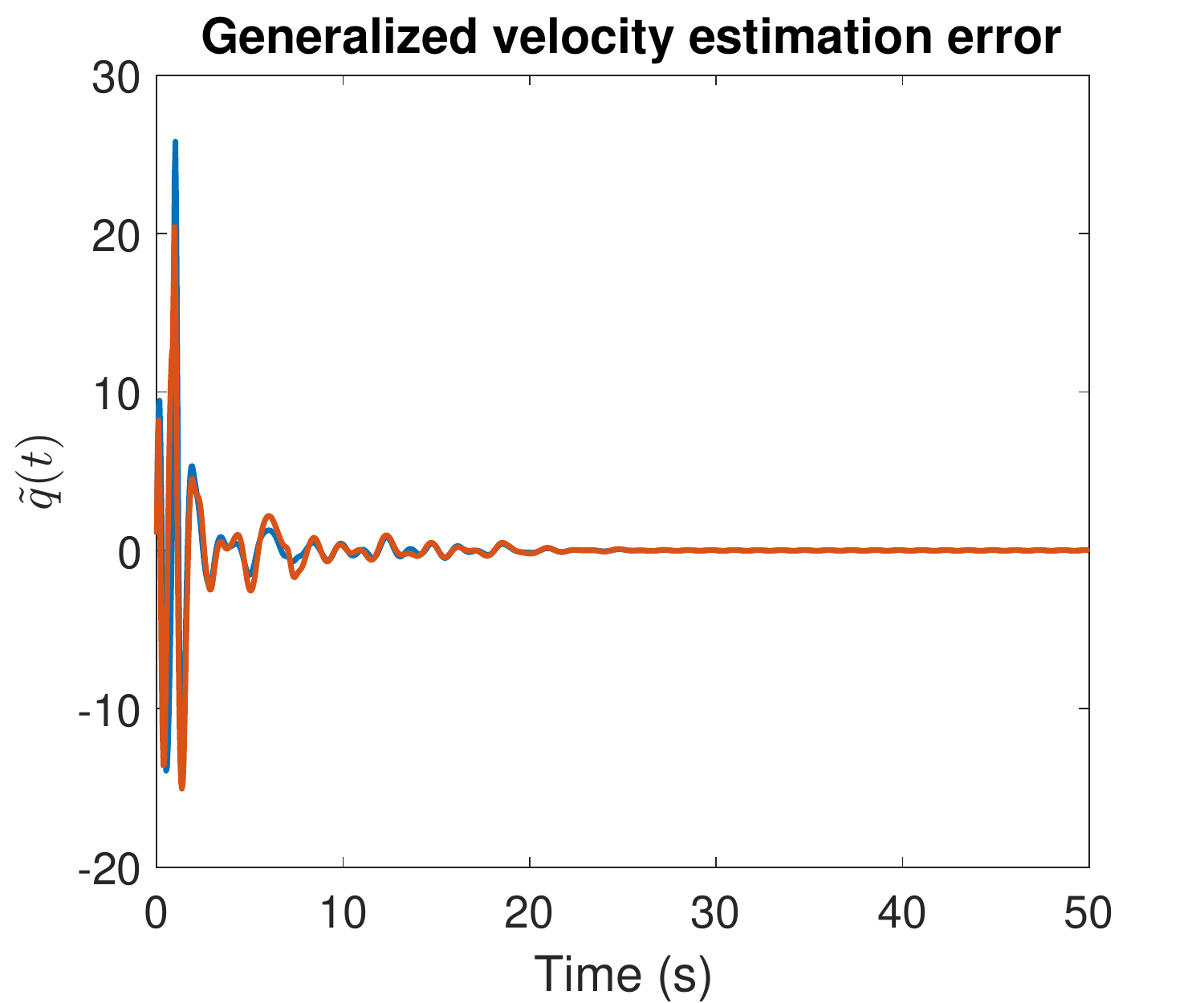}

\caption{\label{fig:qTildeNoNoise}Trajectories of the generalized velocity
estimation errors using noise-free position measurements.}
\end{figure}
\begin{figure}
\includegraphics[width=1\columnwidth]{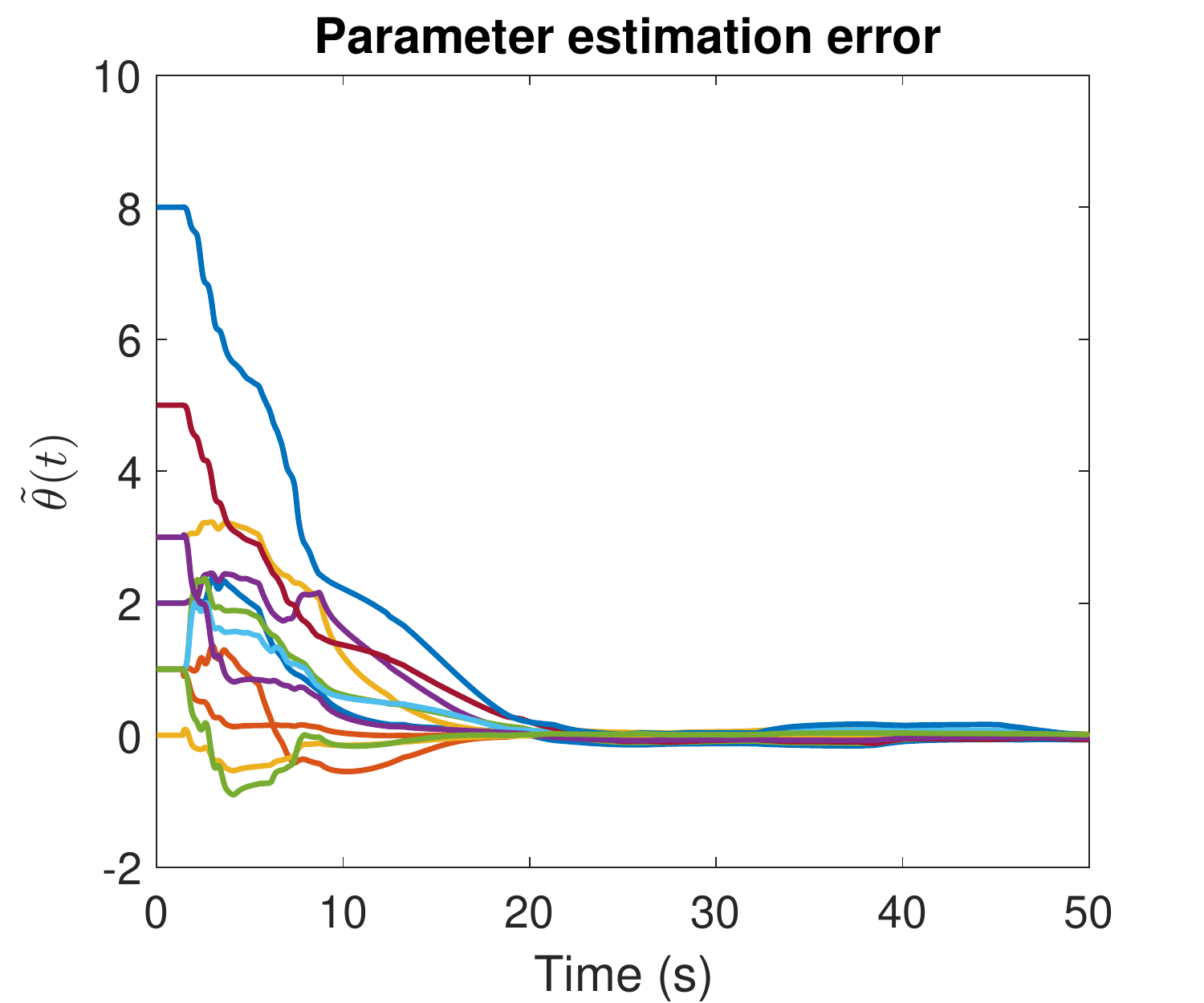}

\caption{\label{fig:ThetaTilde0.001}Trajectories of the parameter estimation
errors with a Gaussian measurement noise (variance = 0.001).}
\end{figure}
\begin{figure}
\includegraphics[width=1\columnwidth]{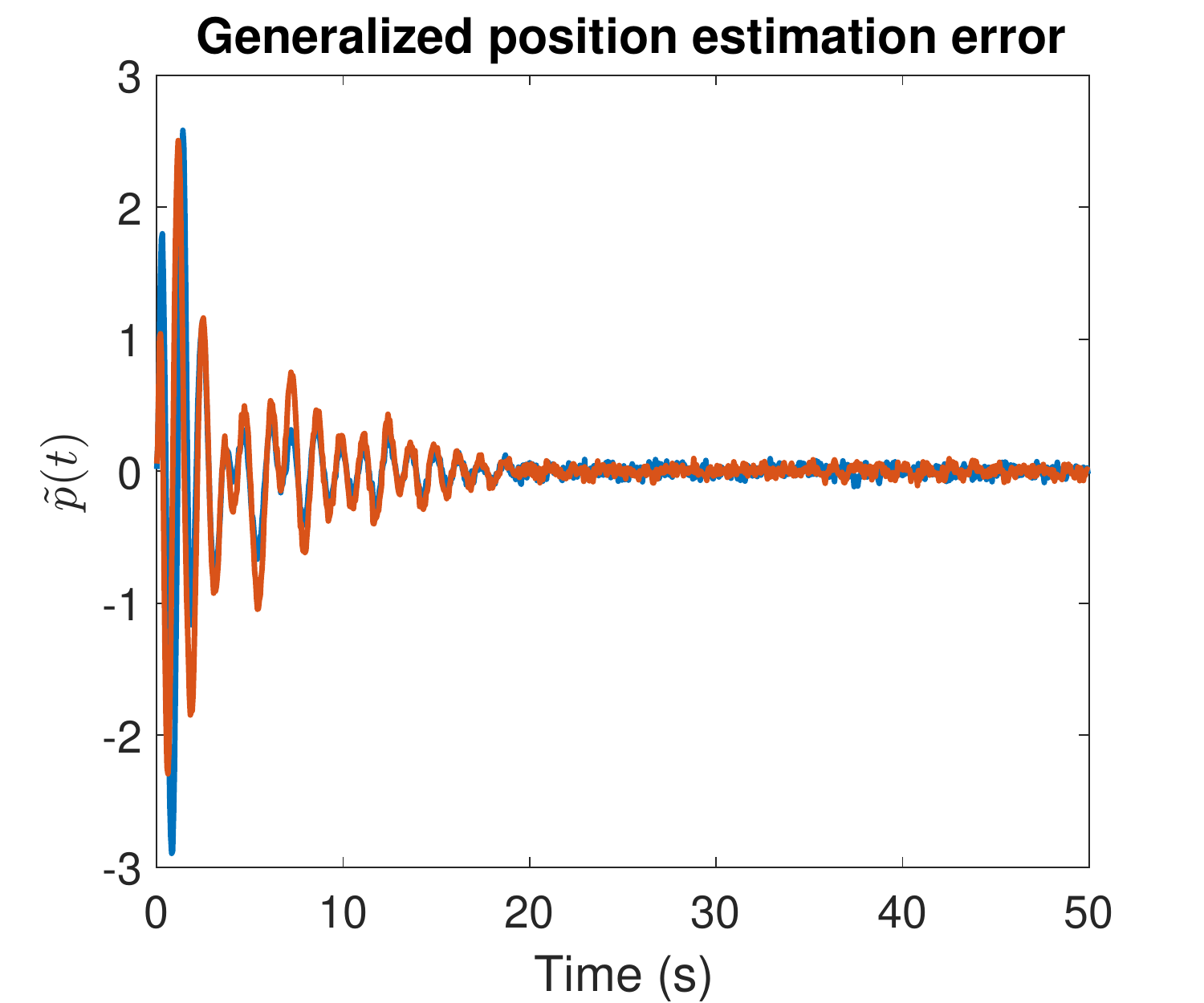}

\caption{\label{fig:pTilde0.001}Trajectories of the generalized position estimation
errors with a Gaussian measurement noise (variance = 0.001).}
\end{figure}
\begin{figure}
\includegraphics[width=1\columnwidth]{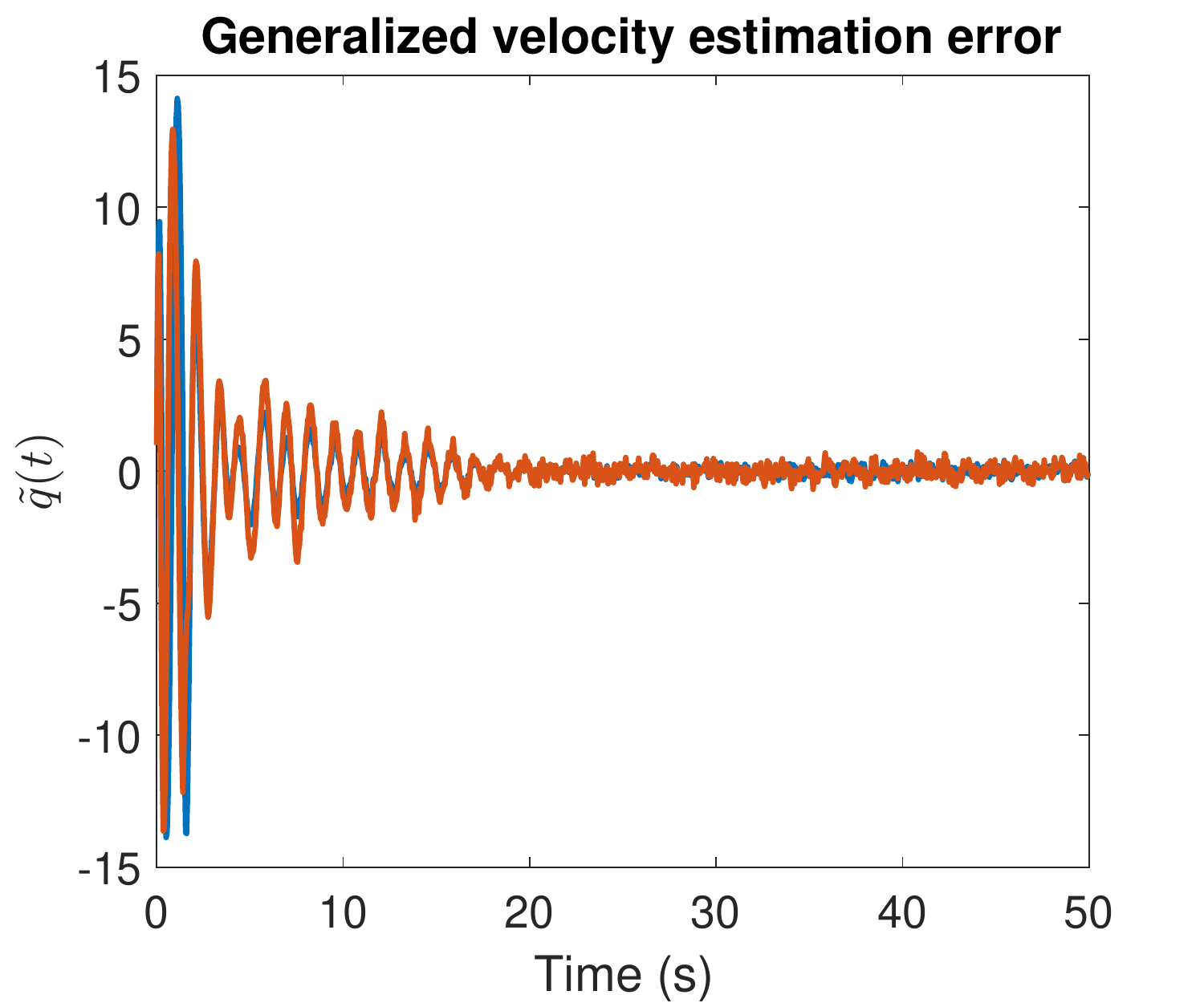}

\caption{\label{fig:qTilde0.001}Trajectories of the generalized velocity estimation
errors with a Gaussian measurement noise (variance = 0.001).}
\end{figure}
\begin{figure}
\includegraphics[width=1\columnwidth]{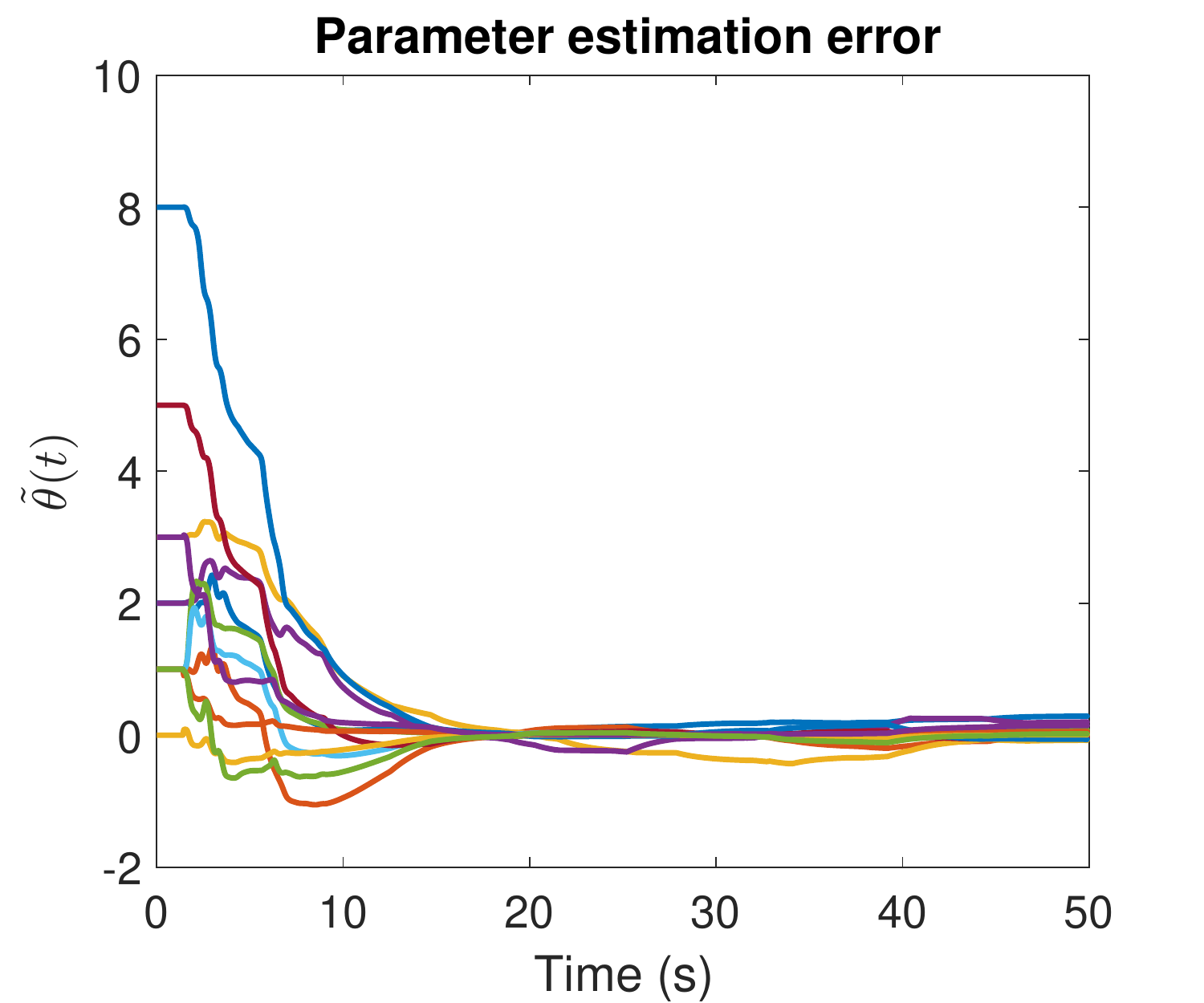}

\caption{\label{fig:ThetaTilde0.01}Trajectories of the parameter estimation
errors with a Gaussian measurement noise (variance = 0.01).}
\end{figure}
\begin{figure}
\includegraphics[width=1\columnwidth]{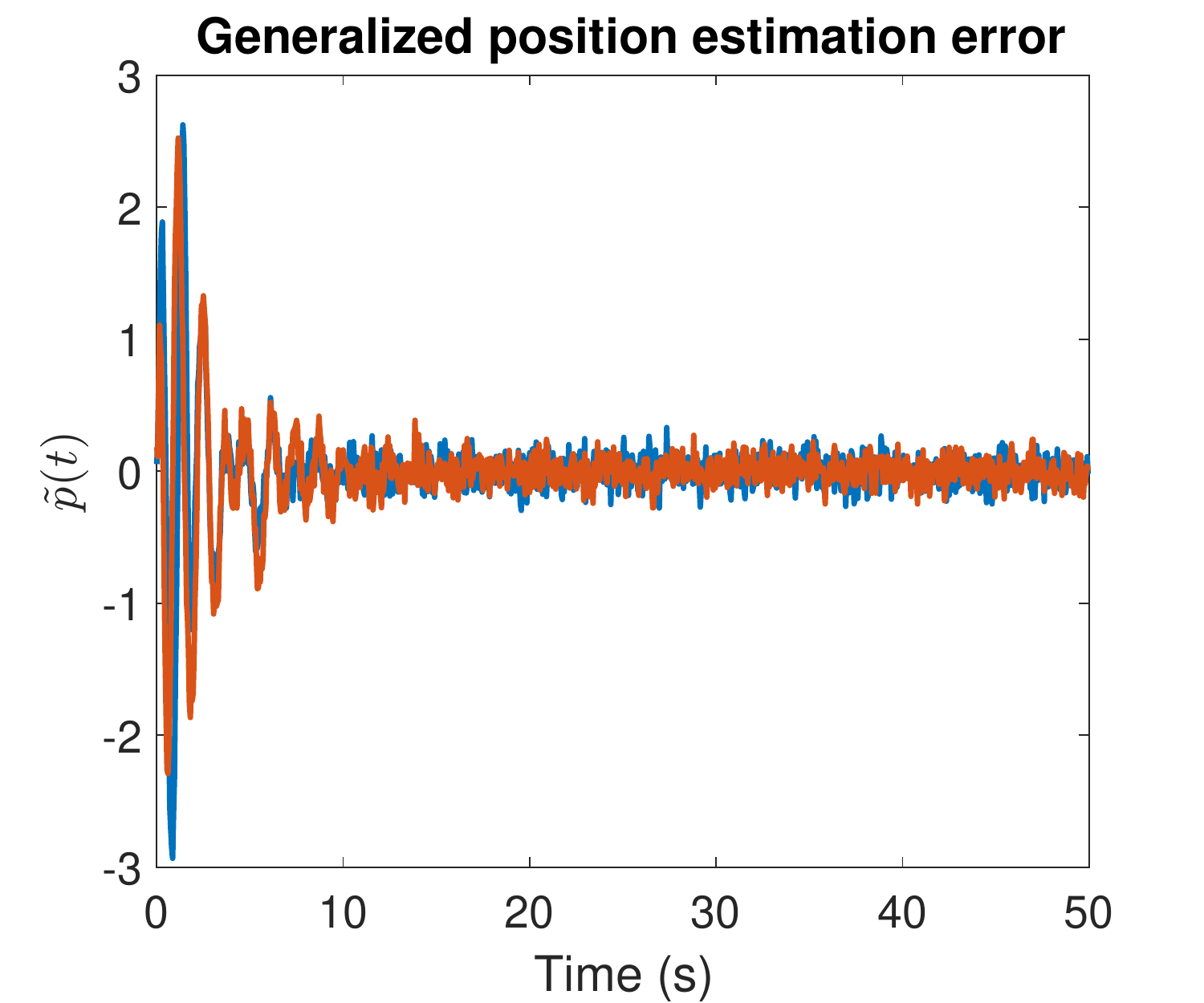}

\caption{\label{fig:pTilde0.01}Trajectories of the generalized position estimation
errors with a Gaussian measurement noise (variance = 0.01).}
\end{figure}
\begin{figure}
\includegraphics[width=1\columnwidth]{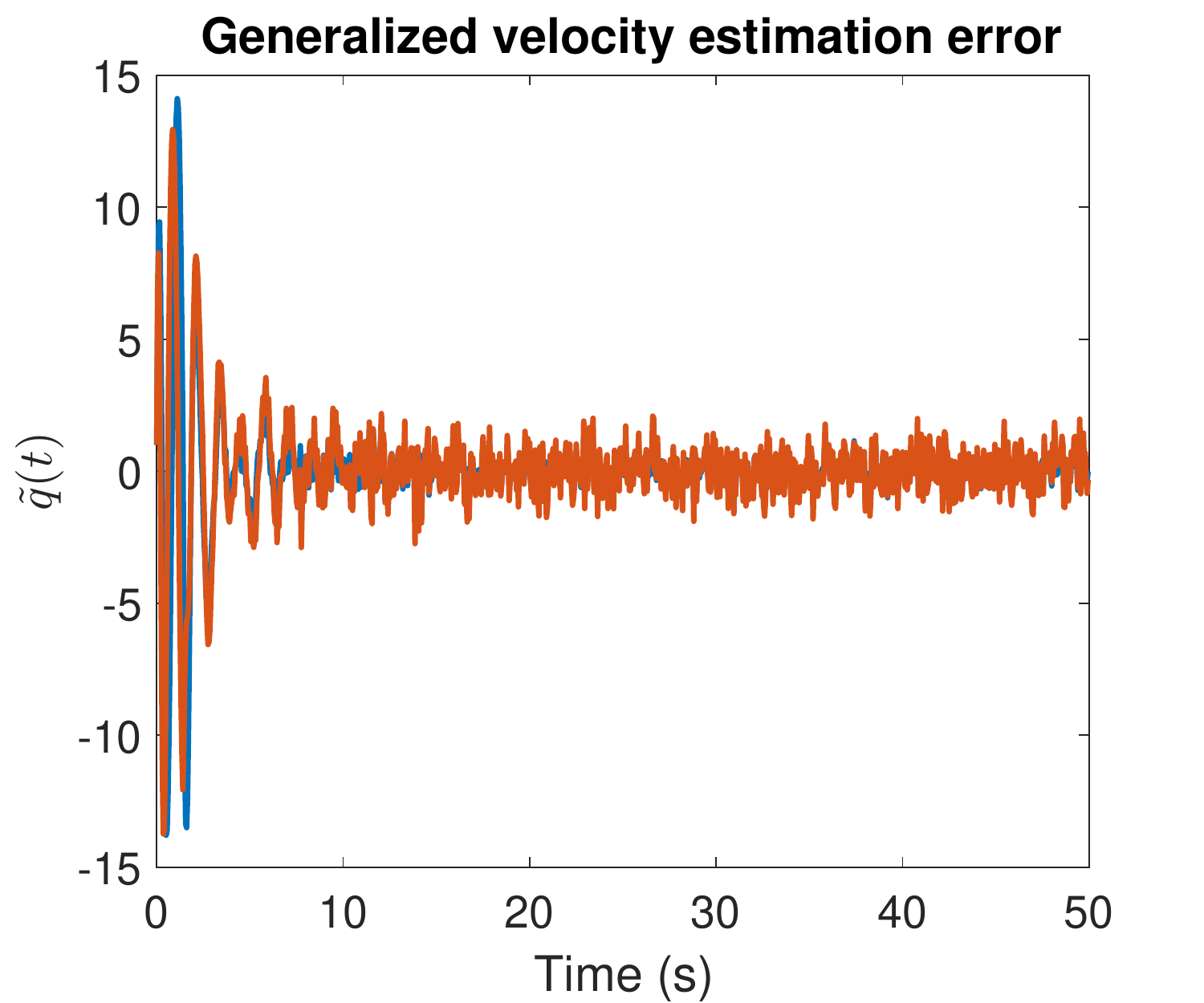}

\caption{\label{fig:qTilde0.01}Trajectories of the generalized velocity estimation
errors with a Gaussian measurement noise (variance = 0.01).}
\end{figure}

The linear system selected for the simulation study is given by
\begin{align*}
\dot{p}\left(t\right) & =q\left(t\right),\\
\dot{q}\left(t\right) & =\begin{bmatrix}2 & 3 & 1 & 5\\
1 & 2 & 1 & 8
\end{bmatrix}\begin{bmatrix}p\left(t\right)\\
q\left(t\right)
\end{bmatrix}+\begin{bmatrix}1 & 3\\
0 & 1
\end{bmatrix}u\left(t\right),
\end{align*}
The contribution of this paper is the design of a parameter estimator
and a velocity observer. The controller is assumed to be any controller
that results in bounded system response. In this simulation study,
the controller, $u$, is designed so that the system tracks the trajectory
$p_{1}\left(t\right)=p_{2}\left(t\right)=\sum_{j=1}^{3}\sin\left(jt\right)+\sin\left(5t\right)$.
Since there are twelve unknown parameters and the desired trajectory
contains only four distinct frequencies, the closed-loop system is
not persistently excited. 

The state observer in (\ref{eq:State observer}) is implemented using
the integral form in (\ref{eq:IntegralUpdate}), and the filter in
(\ref{eq:eta Update}) is implemented using the integral form in (\ref{eq:IntegralUpdateEta}).
The simulation is performed using Euler forward numerical integration
using a sample time of $T_{s}=0.0005$ seconds. Past $\frac{T_{1}+T_{2}}{T_{s}}$
values of the generalized position, $p$, and the control input, $u$,
are stored in a buffer. The matrices $\mathcal{F}$ and $\mathcal{G}$
for the parameter update law in (\ref{eq:Theta Dynamics}) are computed
using trapezoidal integration of the data stored in the aforementioned
buffer. Values of $\mathcal{F}$ and $\mathcal{G}$ are stored in
the history stack and are updated so as to maximize the minimum eigenvalue
of $\mathscr{G}$. 

The initial estimates of the unknown parameters are selected to be
zero, and the history stack is initialized so that all the elements
of the history stack are zero. Data is added to the history stack
using a singular value maximization algorithm. To demonstrate the
utility of the developed method, three simulation runs are performed.
In the first run, the observer is assumed to have access to noise
free measurements of the generalized position. In the second and the
third runs, a zero-mean Gaussian noise with variance 0.001 and 0.01,
respectively, is added to the generalized position signal to simulate
measurement noise. The values of various simulation parameters selected
for the three runs are available in Table \ref{tab:Simulation-parameters-for}.
Figure \ref{fig:ThetaTildeNoNoise} demonstrates that in absence of
noise, the developed parameter estimator drives the parameter estimation
error, $\tilde{\theta}$, to the origin. Figures \ref{fig:pTildeNoNoise}
and \ref{fig:qTildeNoNoise} demonstrates that the developed observer
drives the generalized position and the generalized velocity estimation
error to the origin, respectively. Figures \ref{fig:ThetaTilde0.001}
- \ref{fig:qTilde0.01} indicate that the developed method is applicable
in the presence of measurement noise, with expected degradation of
performance with increasing variance of the noise.

\section{\label{sec:Conclusion}Conclusion}

This paper develops a CL-based adaptive observer and parameter estimator
to estimate the unknown parameter and the generalized velocity of
second-order linear systems using generalized position measurements.
The developed technique utilizes the fact that when integrated twice,
the system dynamics can be reformulated as a set of algebraic equations
that are linear in the unknown parameters. By integrating $n-$times,
the developed method can be generalized to higher-order linear systems.

Simulation results indicate that the developed method is robust to
measurement noise. A theoretical analysis of the developed method
under measurement noise and process noise is a subject for future
research. Future efforts will also focus on the examination the effect
of the integration intervals, $T_{1}$ and $T_{2}$, on the performance
of the observer. 

\bibliographystyle{IEEEtran}
\bibliography{encr,sccmaster,scc}

\begin{thebibliography}{10}
\def\url#1{}
\csname url@samestyle\endcsname
\providecommand{\newblock}{\relax}
\providecommand{\bibinfo}[2]{#2}
\providecommand{\BIBentrySTDinterwordspacing}{\spaceskip=0pt\relax}
\providecommand{\BIBentryALTinterwordstretchfactor}{4}
\providecommand{\BIBentryALTinterwordspacing}{\spaceskip=\fontdimen2\font plus
\BIBentryALTinterwordstretchfactor\fontdimen3\font minus
  \fontdimen4\font\relax}
\providecommand{\BIBforeignlanguage}[2]{{%
\expandafter\ifx\csname l@#1\endcsname\relax
\typeout{** WARNING: IEEEtran.bst: No hyphenation pattern has been}%
\typeout{** loaded for the language `#1'. Using the pattern for}%
\typeout{** the default language instead.}%
\else
\language=\csname l@#1\endcsname
\fi
#2}}
\providecommand{\BIBdecl}{\relax}
\BIBdecl

\bibitem{SCC.Ioannou1996}
P.~Ioannou and J.~Sun, \emph{Robust Adaptive Control}.\hskip 1em plus 0.5em
  minus 0.4em\relax Prentice Hall, 1996.

\bibitem{SCC.Sastry1989a}
S.~Sastry and M.~Bodson, \emph{Adaptive Control: Stability, Convergence, and
  Robustness}.\hskip 1em plus 0.5em minus 0.4em\relax Upper Saddle River, NJ:
  Prentice-Hall, 1989.

\bibitem{SCC.Krstic1995}
M.~Krstic, I.~Kanellakopoulos, and P.~V. Kokotovic, \emph{Nonlinear and
  Adaptive Control Design}.\hskip 1em plus 0.5em minus 0.4em\relax New York,
  NY, USA: John Wiley \& Sons, 1995.

\bibitem{SCC.Duarte.Narendra1989}
M.~A. Duarte and K.~Narendra, ``Combined direct and indirect approach to
  adaptive control,'' \emph{IEEE Trans. Autom. Control}, vol.~34, no.~10, pp.
  1071--1075, Oct 1989.

\bibitem{SCC.Krstic.Kokotovic.ea1993}
M.~Krsti\'{c}, P.~V. Kokotovi\'{c}, and I.~Kanellakopoulos,
  ``Transient-performance improvement with a new class of adaptive
  controllers,'' \emph{Syst. Control Lett.}, vol.~21, no.~6, pp. 451 -- 461,
  1993.

\bibitem{SCC.Chowdhary.Johnson2011a}
G.~V. Chowdhary and E.~N. Johnson, ``Theory and flight-test validation of a
  concurrent-learning adaptive controller,'' \emph{J. Guid. Control Dynam.},
  vol.~34, no.~2, pp. 592--607, Mar. 2011.

\bibitem{SCC.Narendra1987}
K.~S. Narendra and A.~M. Annaswamy, ``A new adaptive law for robust adaptive
  control without persistent excitation,'' \emph{IEEE Trans. Autom. Control},
  vol.~32, pp. 134--145, 1987.

\bibitem{SCC.Narendra.Annaswamy1986}
K.~Narendra and A.~Annaswamy, ``Robust adaptive control in the presence of
  bounded disturbances,'' \emph{IEEE Trans. Autom. Control}, vol.~31, no.~4,
  pp. 306--315, 1986.

\bibitem{SCC.Volyanskyy.Calise.ea2006}
K.~Volyanskyy, A.~Calise, B.-J. Yang, and E.~Lavretsky, ``An error minimization
  method in adaptive control,'' in \emph{Proc. AIAA Guid. Navig. Control
  Conf.}, 2006.

\bibitem{SCC.Chowdhary.Yucelen.ea2012}
G.~Chowdhary, T.~Yucelen, M.~M\"{u}hlegg, and E.~N. Johnson, ``Concurrent
  learning adaptive control of linear systems with exponentially convergent
  bounds,'' \emph{Int. J. Adapt. Control Signal Process.}, vol.~27, no.~4, pp.
  280--301, 2013.

\bibitem{SCC.Kersting.Buss2014}
S.~Kersting and M.~Buss, ``Concurrent learning adaptive identification of
  piecewise affine systems,'' in \emph{IEEE Conf. Decis. Control}, Dec. 2014,
  pp. 3930--3935.

\bibitem{SCC.Chowdhary.Muehlegg.ea2013}
G.~Chowdhary, M.~M\"{u}hlegg, J.~How, and F.~Holzapfel,
  ``\BIBforeignlanguage{English}{Concurrent learning adaptive model predictive
  control},'' in \emph{\BIBforeignlanguage{English}{Advances in Aerospace
  Guidance, Navigation and Control}}, Q.~Chu, B.~Mulder, D.~Choukroun, E.-J.
  van Kampen, C.~de~Visser, and G.~Looye, Eds.\hskip 1em plus 0.5em minus
  0.4em\relax Springer Berlin Heidelberg, 2013, pp. 29--47.

\bibitem{SCC.Modares.Lewis.ea2014}
H.~Modares, F.~L. Lewis, and M.-B. Naghibi-Sistani, ``Integral reinforcement
  learning and experience replay for adaptive optimal control of
  partially-unknown constrained-input continuous-time systems,''
  \emph{Automatica}, vol.~50, no.~1, pp. 193--202, 2014.

\bibitem{SCC.Kamalapurkar.Klotz.ea2014a}
\BIBentryALTinterwordspacing
R.~Kamalapurkar, J.~Klotz, and W.~E. Dixon, ``Concurrent learning-based online
  approximate feedback {N}ash equilibrium solution of {$N$}-player nonzero-sum
  differential games,'' \emph{IEEE/CAA Journal of Automatica Sinica, Special
  Issue on Extensions of Reinforcement Learning and Adaptive Control}, vol.~1,
  no.~3, pp. 239--247, Jul. 2014.
  \url{http://ieeexplore.ieee.org/stamp/stamp.jsp?tp=&arnumber=7004681&isnumber=7004676}
\BIBentrySTDinterwordspacing

\bibitem{SCC.Luo.Wu.ea2014}
B.~Luo, H.-N. Wu, T.~Huang, and D.~Liu, ``Data-based approximate policy
  iteration for affine nonlinear continuous-time optimal control design,''
  \emph{Automatica}, 2014.

\bibitem{SCC.Kamalapurkar.Walters.ea2016}
\BIBentryALTinterwordspacing
R.~Kamalapurkar, P.~Walters, and W.~E. Dixon, ``Model-based reinforcement
  learning for approximate optimal regulation,'' \emph{Automatica}, vol.~64,
  pp. 94--104, Feb. 2016.
  \url{http://www.sciencedirect.com/science/article/pii/S0005109815004392}
\BIBentrySTDinterwordspacing

\bibitem{SCC.Bian.Jiang2016}
T.~Bian and Z.-P. Jiang, ``Value iteration and adaptive dynamic programming for
  data-driven adaptive optimal control design,'' \emph{Automatica}, vol.~71,
  pp. 348--360, 2016.

\bibitem{SCC.Kamalapurkar.Rosenfeld.eatoappear}
R.~Kamalapurkar, J.~A. Rosenfeld, and W.~E. Dixon, ``Efficient model-based
  reinforcement learning for approximate online optimal control,''
  \emph{Automatica}, 2016, to appear.

\bibitem{SCC.DeLaTorre.Chowdhary.ea2013}
G.~D.~L. Torre, G.~Chowdhary, and E.~N. Johnson, ``Concurrent learning adaptive
  control for linear switched systems,'' in \emph{Proc. Am. Control Conf.},
  2013, pp. 854--859.

\bibitem{SCC.Chowdhary.Kingravi.ea2015}
G.~Chowdhary, H.~A. Kingravi, J.~P. How, and P.~A. Vela, ``Bayesian
  nonparametric adaptive control using gaussian processes,'' \emph{IEEE Trans.
  Neural Netw. Learn. Syst.}, vol.~26, no.~3, pp. 537--550, 2015.

\bibitem{arXivKamalapurkar.Reish.ea2015}
R.~Kamalapurkar, B.~Reish, G.~Chowdhary, and W.~E. Dixon. Concurrent learning
  for parameter estimation using dynamic state-derivative estimators.
  {arXiv:1507.08903}.

\bibitem{ParikhKamalapurkarDixonarXiv:1512.03464}
A.~Parikh, R.~Kamalapurkar, and W.~E. Dixon. Integral concurrent learning:
  Adaptive control with parameter convergence without {PE} or state
  derivatives. {arXiv}:1512.03464.

\bibitem{SCC.Chowdhary2010}
G.~Chowdhary and E.~Johnson, ``Concurrent learning for convergence in adaptive
  control without persistency of excitation,'' in \emph{IEEE Conf. Decis.
  Control}, 2010, pp. 3674--3679.

\bibitem{SCC.Xian2004c}
B.~Xian, M.~S. de~Queiroz, D.~M. Dawson, and M.~McIntyre, ``A discontinuous
  output feedback controller and velocity observer for nonlinear mechanical
  systems,'' \emph{Automatica}, vol.~40, no.~4, pp. 695--700, 2004.

\bibitem{SCC.Khalil2002}
H.~K. Khalil, \emph{Nonlinear Systems}, 3rd~ed.\hskip 1em plus 0.5em minus
  0.4em\relax Upper Saddle River, NJ: Prentice Hall, 2002.

\end{thebibliography}

\end{document}